\documentclass[10pt, conf, twocolumn]{IEEEtran}
\usepackage{color}
\usepackage{times}
\usepackage{epsfig}
\usepackage{graphicx}
\usepackage{epstopdf}
\usepackage{algorithm}
\usepackage{algorithmic}
\usepackage{amsmath}
\usepackage{amssymb}
\usepackage{amsxtra}
\usepackage{multirow}
\usepackage{mathtools}
\usepackage{caption}
\usepackage{cleveref}
\usepackage{url}
\usepackage{subfigure}
\usepackage{cite}
\usepackage{subfigure}
\usepackage{amsthm}
\usepackage{cleveref}

\newtheorem{lemma}{Lemma}
\newtheorem{proposition}{Proposition}
\newtheorem{corollary}{Corollary}

\newtheorem*{proof*}{Proof}

\theoremstyle{definition}
\newtheorem{definition}{Definition}[]

\begin{document}

\title{Optimal Dynamic Contract for Spectrum Reservation in Mission-Critical UNB-IoT Systems}

\author{ \IEEEauthorblockN{\large Muhammad Junaid Farooq and Quanyan Zhu} \\ \IEEEauthorblockA{Department of Electrical \& Computer Engineering, Tandon School of Engineering, \\New York University, Brooklyn, NY, USA,} Emails: \{mjf514, qz494\}@nyu.edu. \vspace{-0.0in}
}

\maketitle

\begin{abstract}
Spectrum reservation is emerging as one of the potential solutions to cater for the communication needs of massive number of wireless Internet of Things (IoT) devices with reliability constraints particularly in mission-critical scenarios. In most mission-critical systems, the true utility of a reservation may not be completely known ahead of time as the unforseen events might not be completely predictable. In this paper, we present a dynamic contract approach where an advance payment is made at the time of reservation based on partial information about spectrum reservation utility. Once the complete information is obtained, a rebate on the payment is made if the reservation is released. In this paper, we present a contract theoretic approach to design an incentivized mechanism that coerces the applications to reveal their true application type resulting in greater profitability of the IoT network operator. The operator offers a menu of contracts with advanced payments and rebate to the IoT applications without having knowledge about the types of applications. The decision of the applications in selecting a contract leads to a revelation of their true type to the operator which allows it to generate higher profits than a traditional spectrum auction mechanism. Under some assumptions on distribution of the utility of the applications, closed form solutions for the optimal dynamic spectrum reservation contract are provided and the sensitivity against system parameters is analyzed.
\end{abstract}

\IEEEpeerreviewmaketitle


\vspace{-0.0in}
\section{Introduction}

The Internet of things (IoT) is foreseen to revolutionize the operations, management, and control of electronic systems around us. Most existing IoT devices utilize traditional wireless personal area network (WPAN) communication technologies~\cite{protocols} such as WiFi, Bluetooth, Zigbee, etc., which are inherently short-range privately administered networks. However, the current focus of network operators is towards developing low-power wide-area network (LPWA) technologies~\cite{IoT_access} dedicated for IoT communication for providing reliable communication and services for large scale IoT systems in smart cities along the same lines as the cellular data networks.

Due to a predicted massive surge in the number of IoT devices in the future owing to widespread adoption of the technology~\cite{junaid_iobt,junaid_acc}, there will be an acute shortage of wireless spectrum for dedicated allocation to these systems. This will pose a challenge to the mission-critical (MC) IoT systems~\cite{mc_definition} such as in public safety systems or other emergency networks~\cite{junaid_globecom} requiring dedicated spectrum availability at all times due to the unpredictability of unforseen events. The MC applications~\cite{MC_5G} may be highly delay-sensitive, e.g., real time systems involving artificial intelligence (AI), virtual and augmented reality (VR/AR), real-time control loops, streaming analytics, etc. Such applications are termed as MC not only due to the link with life-threatening situations but also due to a substantial risk of malfunction, services interruption, and enterprise operation jeopardy resulting in damages and losses to property, individuals, or businesses, etc. In MC applications, often a delay in communication may in fact fail the initial objective of the application. For instance, in a surveillance system where an unusual activity needs to be reported promptly to successfully avoid any potential damage or loss of property and a report beyond a certain delay may be futile.

The spectrum requirements for massive IoT networks~\cite{massive_iot} will exceed the capacity of the unlicensed spectrum bands. Therefore, exploring the sub-GHz spectrum and opportunistically accessing whitespaces in existing licensed spectrum bands will be inevitable to cater for the massive wireless connectivity demand. Several traditional approaches have been developed for mitigating this issue such as new MAC layer schemes~\cite{iot_mac} to improve the efficiency of multiple access as well as cognitive radio technologies (CRN)~\cite{novel_reservation_mac_cognitive_radio} for opportunistically accessing licensed spectra. One of the candidate LPWA technologies that has emerged recently is known as ultra narrow band (UNB)~\cite{unb_report} which uses the random frequency and time multiple access (RFTMA)~\cite{time-frequency} at the MAC layer in which each user selects a time and frequency randomly for transmission. Although it can improve the capacity of the system, however, there is no performance and reliability guarantees which may be crucial for mission critical applications.

\begin{figure}
  \centering
  \includegraphics[width=3.2in]{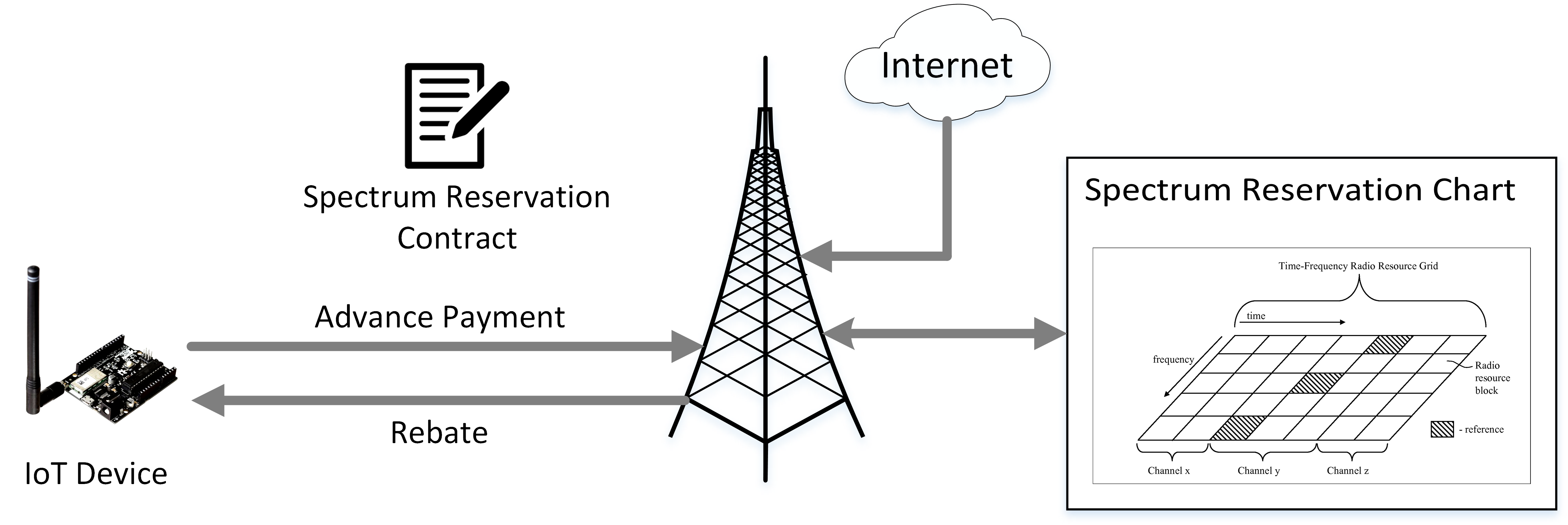}\\
  \caption{A spectrum reservation based UNB-IoT system where the IoT devices only communicate with the AP during the designated time-frequency blocks. The AP offers a dynamic contract with advance payment and rebate amounts to the devices and maintains a spectrum reservation chart for a fixed time period.\vspace{-0.0in}}\label{Fig:sys_model}
\end{figure}

Spectrum reservation is an essential approach to provide guarantees for MC applications. It has been shown that reservation of spectrum leads to efficiency and reliability of communication which is particularly useful in MC and emergency applications~\cite{slot_reservation}. However, the implementation of spectrum reservation needs to be appropriately incentivized for it to be used in practice. Most works in literature dealing with spectrum reservation are focused mainly on the protocol design aspects such as~\cite{access_reservation_protocol,periodic_reservation}. Those that deal with the economic perspectives only investigate static contracts to establish the quantity and price of spectrum to be reserved which assume perfect knowledge of the application requirements~\cite{TV_white_space}. However, in most practical situations particularly emergency networks, the need for spectrum access cannot be perfectly known ahead of time. If the spectrum has not been reserved \emph{a priori}, then the application may have to contend for channel access resulting in significant delays in the communication which might be costly in MC and emergency scenarios. One of the solutions to this problem is that the network operators offer dynamic contracts to applications whereby an advanced payment is made earlier to reserve the spectrum in the form of a time-frequency (TF) block for future use. However, prior to actually using the spectrum, the application may request for a rebate and release the reserved spectrum for use by other applications. Such dynamic contracts can foresee the risks and uncertainties in the future and reduce them through reservation and prioritization.
\begin{figure}[t]
  \centering
  \includegraphics[width=3.5in]{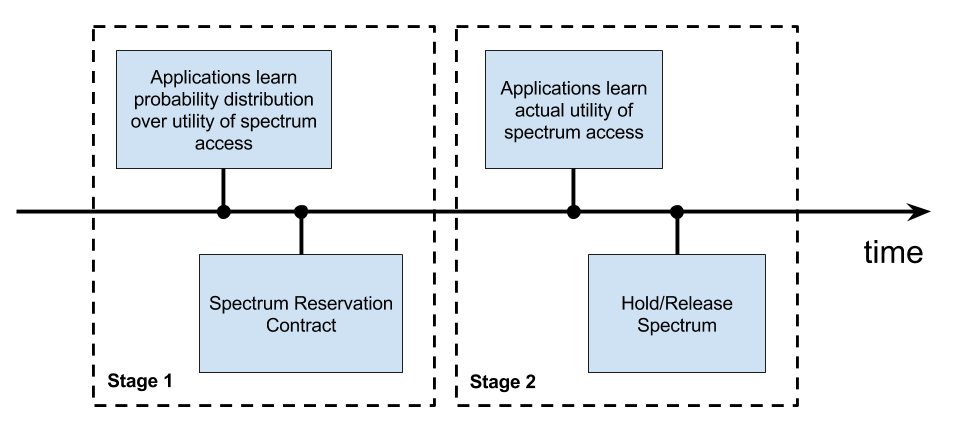}\\
  \caption{Stages of spectrum reservation contract. In the first stage, the applications privately obtain a probability distribution over their utility of the spectrum reservation based on which they opt for one of the contracts by the operator. Once the true utility of the reserved spectrum is known, the applications either hold the spectrum or release it for the agreed rebate amount.\vspace{-0.0in}}\label{Fig:contract}
\end{figure}

In this paper, we develop a dynamic mechanism for spectrum reservation considering the uncertainty in available spectrum at each time and the uncertainty in the requirement for spectrum access. We make use of tools from sequential screening~\cite{sequential_screening} and mechanism design literature to establish a dynamic menu of contracts which comprises of an advanced payment for spectrum reservation in the future along with a rebate policy if the spectrum is released before the time of spectrum access. This allows the network operator to discriminate the unknown application types and generate higher profits than the traditional auction mechanisms where every application is completely aware of its true utility. We assume a two-type categorization of IoT applications where they are either classified as MC or non-MC and consequently an optimal binary contract is designed by the service provider. Based on assumptions on the distribution of utility of the MC and non-MC applications, closed form results for the optimal contracts are derived and the effect of system parameters is analyzed to gain insights.


\vspace{-0.0in}
\section{System Model}
In this section, we first describe the network model comprising of details regarding the availability and cost of obtaining an idle TF block. Then, a description of utility achieved by applications from reserving the spectrum in advance is discussed.

\begin{figure}[t]
  \centering
  \includegraphics[width=3.0in]{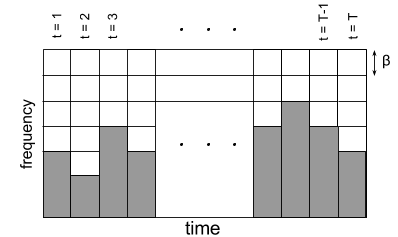}\\
  \caption{Available spectrum divided into time-frequency blocks for reservation. The shaded regions indicate blocks occupied by the primary user.\vspace{-0.0in}}\label{Fig:TF}
\end{figure}

\vspace{-0.0in}
\subsection{Network Model}
We assume that a single IoT operator is coordinating the communication between low power IoT devices using UNB transmissions. We consider a single access point (AP) serving a wide area network of IoT devices. The AP reserves TF blocks in the available whitespace in existing licensed spectra for a fixed duration $T$ in the future. Let $n_t$ denote the number of available channels of equal bandwidth $\beta$ at time $t$. An illustration of the spectrum resources available to the AP for the duration $T$ is provided in Fig.~\ref{Fig:TF}. At each time $t = 1, \ldots, T$, a random number of channels is available due to the uncontrolled activity of the primary users in licensed bands.
Let $\kappa_t = g(n_t)$, where $g: \mathbb{Z}^+ \rightarrow \mathbb{R}$, represents the cost of obtaining a channel at time $t$ that is related to the number of available channels. We denote the time average of the channel cost over the duration $T$ by $\kappa = \frac{1}{T}\sum_{t=1}^{T} \kappa_t$.



The IoT applications are categorized into two broad types, i.e., MC or non-MC. Let $\pi_c \in [0,1]$ represent the proportion of MC applications and $\pi_n = 1 - \pi_c$ be the proportion of non-MC IoT applications. The AP enforces a reservation based access scheme in which the applications reserve TF blocks for the duration $T$ in advance. If before the time of channel access, the application determines that the transmission is not needed, it can release the TF block in advance and request for a partial rebate on the initial payment\footnote{The payments can be realized in several ways such using credits where each device has a fixed number of credits to use for spectrum access at the beginning of the first stage.}. Hence, this transaction can be summarized into two stages. At the beginning of the first stage, the applications privately make an assessment of their transmission requirements in the future. Since emergency situations and or other unforseen events cannot be completely predicted, the applications obtain a probability distribution over their utility of reserving the channel. At the end of the first stage, the service provider and the applications enter into a contract which includes an advance payment and a rebate agreement. At the beginning of the second period, the applications learn their true utility of the reserved spectrum. If the utility is higher than the rebated amount in monetary terms, the application holds the channel. Otherwise, it prefers to release the channel and obtain a rebate at the end of the second period.




For each time instant, $t= 1, \ldots, T$, two contracts are offered by the network operator based on the average cost of obtaining an idle channel. The proportion of MC and non-MC applications vary during each time slot due to the relative frequency of transmissions of each type of application. For instance, if a MC application has has to transmit at each time slot, then the number of contending MC applications will be the same for all time slots. Similarly, if the non-MC applications have to transmit once after $10$ time slots, then the non-MC applications will only contribute in the proportion once after every $10$ time slots.

\subsection{Utility of Spectrum Reservation}
In the absence of spectrum reservation, the IoT devices have to contend for spectrum access by employing a multiple access protocol involving some randomness in which case, the average delay in channel access or the average number of transmission until successful message delivery is denoted by $\delta$. It is natural to associate the spectrum reservation utility of the applications to the expected channel access delay experienced if there was no reserved spectrum. The average access delay is a function of the number of simultaneously transmitting devices, their physical placement relative to other neighbouring devices, and the multiple access protocol used.
Assuming that $\delta$ represents the waiting time until successful spectrum access, we model it as an exponential random variable with mean $\frac{1}{\lambda}$. Let the utility of an application for reserving a TF block in the future be represented by $V_{\chi} = h_{\chi}(\delta), \chi \in \{c,n\}$ referring to the MC and non-MC applications respectively. $h_{\chi}(.)$ is a transformation of the average delay into a monetary value. We assume that the transformation $h(.)$ does not change the distribution of the delay except the mean and the variance. Therefore, we can claim that $V_{\chi}$ is also exponentially distributed with intensity $\frac{1}{\lambda_{\chi}}, \chi \in \{\text{c,n}\}$. Consequently, we can state that $f_{V_{\chi}}(v) = \lambda_{\chi} e^{-\lambda_{\chi} v}$ and $F_{V_{\chi}}(v) = 1 - e^{-\lambda_{\chi}v}$. With a slight abuse of notation, we will refer to these functions as simply $f_{\chi}(v)$ and $F_{\chi}(v), \chi \in \{\text{c,n}\}$ in the sequel. Furthermore, we assume that $\lambda_{\text{c}}  \leq \lambda_{\text{n}}$, which implies that the MC applications have a higher utility of reserving the channel as compared to the non-MC applications. This is a reasonable assumption as, inherently, MC applications are more sensitive to the average delay in spectrum access.

\vspace{-0.0in}
\section{Methodology}
In this section, we first describe the objective of the IoT service provider and then provide a detailed description of the methodology used to achieve the desired objective.

\subsection{Objective}
The objective of the IoT network operator is to create a menu of contracts for the two main types of IoT applications which allows it to
coerce them into revealing their private information and using it to generate higher profits. The challenge lies in the information asymmetry between the applications and the service provider. The applications know more about their utility of a TF block than the service provider. If the service provider had information about the type of application requesting the TF blocks, it can use price discrimination to extract maximum profit by charging higher prices to MC applications. However, in the absence of information about the applications, the service provider has to design a mechanism which results in the applications revealing their true types by selecting the contract that suits them. This mechanism is in the form of a refund contract where the applications learn their utilities sequentially.


In summary, the objective of the operator is to optimally design the advance payments and rebates for the MC and non-MC applications represented by the tuple $\mathcal{C} = \{p_\text{c}, r_{\text{c}}, p_{\text{n}}, r_{\text{n}}\}$, where $p_{\text{c}}$ and $p_{\text{n}}$ are the advance payments respectively at the end of the first stage stage while $r_{\text{c}}$ and $r_{\text{n}}$ is the rebate offered by the network operator in case the applications release the reserved TF blocks at the end of the second stage. The IoT applications will select one of the contracts and if the mechanism is well designed, the applications will not have any incentive to deviate from their true preferences. Hence, in this process, they reveal their true types to the operator leading to higher profitability.

\subsection{Spectrum Reservation Contract}
The contract is established at the end of the first stage while only a probability distribution over the applications' valuation is known to them privately.


\subsubsection{Operator Profitability}
Given the contract tuple $\mathcal{C} = \{p_c, r_c, p_n, r_n\}$, the expected profit of the IoT network operator can be expressed as follows:
\begin{align}
\Pi(\mathcal{C}) &= \sum_{\chi \in \{\text{c},\text{n}\}} \pi_\chi \left( p_{\chi} - r_{\chi} F_{\chi}(r_{\chi})  - \kappa(1 - F_{\chi}(r_{\chi}))\right), \notag \\
&= \sum_{\chi \in \{\text{c},\text{n}\}} \pi_\chi \left( p_{\chi} - r_{\chi} +    e^{-\lambda_{\chi}r_{\chi}} (r_{\chi} - \kappa) \right).
\end{align}
Note that $p_{\chi}$ is the advance payment received by the network operator, $r_{\chi} F_{\chi}(r_{\chi})$ is the expected amount of rebate paid back to the application, and $\kappa(1 - F_{\chi}(r_{\chi}))$ is the expected cost of obtaining an idle channel by the operator. The expected profit from both MC and non-MC applications are added after being weighted by their proportions in the network.

\subsubsection{IC \& IR Constraints}
In order for the spectrum reservation mechanism to be implementable, it has to be individually rational (IR), i.e., both the MC and non-MC applications will only participate in the spectrum reservation if at best on average, they do not receive a lower utility than the payment they make. Note that at the end of the second stage, the application will always prefer to release the channel in return for a rebate if the amount refunded is greater that its actual utility of the channel. Therefore, the expected return of the application can be written as $\int_{0}^{\infty} \mathcal{R}_{\chi}(v) f_{\chi}(v)dv$, where $\mathcal{R}_{\chi}(v) = \max(r_{\chi},v)$.
Hence, the expected utility of the application for reserving the channel can be expressed as $r_{\chi}F_{\chi}(r_{\chi}) + \int_{r_{\chi}}^{\infty} x f_X(x) dx$ and the IR implies that it should be at least as much as the payment made for reservation. This can be expressed as follows:
\begin{align}
r_{\chi}F_{\chi}(r_{\chi}) + \int_{r_{\chi}}^{\infty} x f_X(x) dx \geq p_{\chi}, \ \chi \in \{\text{c},\text{n}\}.
\end{align}
Moreover, the mechanism also needs to be incentive compatible (IC), i.e., there is no incentive for any type of application to hide their true type from the operator by choosing a different contract. In other words, the MC applications should be better off choosing the MC contract and vice versa. For the two-type case, this can be formally expressed by the following two constraints:
\begin{align}
r_{\text{c}} F_{\text{c}}(r_{\text{c}}) + \hspace{-0.1cm} \int_{r_{\text{c}}}^{\infty} \hspace{-0.2cm} \nu f_{\text{c}}(\nu) d\nu - p_{\text{c}} \geq r_{\text{n}} F_{\text{c}}(r_{\text{n}}) + \hspace{-0.1cm} \int_{r_{\text{n}}}^{\infty} \hspace{-0.2cm} \nu f_{\text{c}}(\nu) d\nu - p_{\text{n}},
\end{align}
\begin{align}
r_{\text{n}} F_{\text{n}}(r_{\text{n}}) + \hspace{-0.1cm} \int_{r_{\text{n}}}^{\infty} \hspace{-0.2cm} \nu f_{\text{n}}(\nu) d\nu - p_{\text{n}} \geq r_{\text{c}} F_{\text{n}}(r_{\text{c}}) + \hspace{-0.1cm} \int_{r_{\text{c}}}^{\infty} \hspace{-0.2cm} \nu f_{\text{n}}(\nu) d\nu - p_{\text{c}},
\end{align}
The next step is combine these set of equations for the case considered to write the optimal contracting problem.
\subsubsection{Optimal Contracting Problem}
The optimal contracting problem can then be formally written as follows:
\begin{align}
&\underset{\mathcal{C}}{\max}
& & \ \hspace{-0.4in} \Pi(\mathcal{C})= \sum_{\chi \in \{\text{c},\text{n}\}} \pi_\chi \left( p_{\chi} - r_{\chi} +    e^{-\lambda_{\chi}r_{\chi}} (r_{\chi} - \kappa) \right), \label{opt11}\\
& \text{subject to}
& & \notag\\
& \text{(IR$_{\chi})$}
& & r_{\chi} + \frac{1}{\lambda_{\chi}}e^{-\lambda_{\chi}r_{\chi}} \geq p_{\chi} , \; \chi = \{\text{c}, \text{n}\}, \label{IR_consts}\\
& \text{(IC$_{\text{c,n}})$}
& & r_{\text{c}} + \frac{e^{-\lambda_{\text{c}} r_{\text{c}}}}{\lambda_{\text{c}}} - p_{\text{c}} \geq r_{\text{n}} + \frac{e^{-\lambda_{\text{c}} r_{\text{n}}}}{\lambda_{\text{c}}} - p_{\text{n}}, \label{IC_const_c}\\
& \text{(IC$_{\text{n,c}})$}
& & r_{\text{n}} + \frac{e^{-\lambda_{\text{n}} r_{\text{n}}}}{\lambda_{\text{n}}} - p_{\text{n}} \geq r_{\text{c}} + \frac{e^{-\lambda_{\text{n}} r_{\text{c}}}}{\lambda_{\text{n}}} - p_{\text{c}}. \label{IC_const_n}
\end{align}
The solution to this problem results in the optimal contract $\mathcal{C}$ that leads to the maximum profit for the network operator in the scenario where the MC applications learn their utility of channel reservation over time in multiple stages.

\subsection{Solution to the Optimization Problem}
Before we proceed towards the solution to the problem, we first exploit the structure of the constraints to remove the redundancies leading to a simplification of the original problem. The special structure emerges due to the properties of the exponentially distributed utilities of the applications. We describe the notion of first-order stochastic dominance (FSD) in the following definition.
\begin{definition}\label{FSD_definiton}
The distribution of a random variable $X \in \mathcal{X}$ first order stochastically dominates the distribution of a random variable $Y \in \mathcal{X}$ if $F_X(x) \leq F_{Y}(x), \forall x \in \mathcal{X}$.
\end{definition}

According to the definition, it is clear that the distribution of utility of MC applications $F_{\text{c}}(v)$ first order stochastically dominates the distribution of non-MC applications $F_{\text{n}}(v)$ since $\lambda_c \leq \lambda_n$. The implication of this property on the original optimization problem described by \cref{opt11,IR_consts,IC_const_c,IC_const_n} is provided by the following proposition.


\begin{proposition}
Under the FSD of the distribution of utility of MC applications, the constraints IC$_{\text{c,n}}$, and IR$_{\text{n}}$ imply IR$_{\text{c}}$. Hence, the constraint IR$_{\text{c}}$ is redundant and can be removed from the problem.
\begin{proof}
The constraints IC$_{\text{c,n}}$ and IR$_{\text{n}}$ are expressed as follows:
\begin{align}
r_{\text{c}} + \frac{e^{-\lambda_{\text{c}} r_{\text{c}}}}{\lambda_{\text{c}}} - p_{\text{c}} \geq r_{\text{n}} + \frac{e^{-\lambda_{\text{c}} r_{\text{n}}}}{\lambda_{\text{c}}} - p_{\text{n}},
\end{align}
\begin{align}
r_{\text{n}} + \frac{1}{\lambda_{\text{n}}}e^{-\lambda_{\text{n}}r_{\text{n}}} - p_{\text{n}}\geq 0.
\end{align}
Since $\lambda_{\text{c}} \leq \lambda_{\text{n}}$, so it clear that $r_{\text{n}} + \frac{e^{-\lambda_{\text{c}} r_{\text{n}}}}{\lambda_{\text{c}}} - p_{\text{n}}  \geq r_{\text{n}} + \frac{1}{\lambda_{\text{n}}}e^{-\lambda_{\text{n}}r_{\text{n}}} - p_{\text{n}} \geq 0$. This in turn implies that $r_{\text{c}} + \frac{e^{-\lambda_{\text{c}} r_{\text{c}}}}{\lambda_{\text{c}}} - p_{\text{c}} \geq 0$, which precisely describes the constraint IR$_{\text{c}}$. Hence, we can remove it from the original problem.
\end{proof}
\end{proposition}

Furthermore, note that in the optimal contract, the constraint IR$_{\text{n}}$ is binding, i.e., satisfied with equality. If it does not bind, then increasing $p_{\chi}, \chi \in \{\text{c,n}\}$ equally can lead to an increase in the profit of the IoT network operator. Similarly, the constraint IC$_{\text{c,n}}$ also binds since otherwise increasing $p_{c}$ can lead to an increase in profit. Therefore, we can substitute the constraints IR$_{\text{n}}$ and IC$_{\text{c,n}}$ into the objective function and ignore IC$_{\text{n,c}}$ to obtain a relaxed problem as follows:
\begin{align}
&{ \underset{r_{\text{c}}, r_{\text{n}}}{\max} }
& & e^{-\lambda_{\text{n}} r_{\text{n}}} \left( \frac{1}{\lambda_{\text{n}}}  +  \pi_{\text{n}} (r_{\text{n}} - \kappa) \right) - e^{-\lambda_{\text{c}} r_{\text{n}}} \left( \frac{\pi_{\text{c}}}{\lambda_{\text{c}}} \right) + \notag \\
&&& \pi_{\text{c}} \left( e^{-\lambda_{\text{c}} r_{\text{c}}} \left( \frac{1}{\lambda_{\text{c}}} + r_{\text{c}} - \kappa \right)  \right).
\end{align}
Fortunately, the relaxed problem is separable in the decision variables $r_{\text{c}}$ and $r_{\text{n}}$. Hence, the optimal solution to the relaxed problem can be expressed by the following lemma.

\begin{lemma}\label{rebate_lemma}
In the optimal contract offered to the IoT applications, the rebate to the MC applications that maximizes the expected profit equals the average cost of the channel, i.e., $r_{\text{c}}^* = \kappa$. The optimal rebate for the non-MC applications can be obtained by solving the following fixed-point equation:
\begin{align}\label{fixed_point}
r_{\text{n}}^* = \kappa + \frac{ \pi_{\text{c}} ( e^{ r_{\text{n}}^*(\lambda_{\text{n}} - \lambda_{\text{c}})}   - 1   ) }{\pi_{\text{n}} \lambda_{\text{n}}}
\end{align}
\begin{proof}
Let $J(r_{\text{c}}) = \pi_{\text{c}} \left( e^{-\lambda_{\text{c}} r_{\text{c}}} \left( \frac{1}{\lambda_{\text{c}}} + r_{\text{c}} - \kappa \right)  \right)$. Since $\frac{dJ(r_{\text{c}})}{dr_{\text{c}}} = - \lambda_{\text{c}} (r_{\text{c}} - \kappa)$, the optimal amount of rebate offered to the MC applications $r_{\text{c}}^* = \kappa$ and it is indeed a maximizer as $\frac{d^{2} J(r_{\text{c}})}{dr_{\text{c}}^2} = - \lambda_{\text{c}} < 0$.
Similarly, let $H(r_{\text{n}}) = e^{-\lambda_{\text{n}} r_{\text{n}}} \left( \frac{\pi_{\text{c}} + \pi_{\text{n}}}{\lambda_{\text{n}}}  +  \pi_{\text{n}} (r_{\text{n}} - \kappa) \right) - e^{-\lambda_{\text{c}} r_{\text{n}}} \left( \frac{\pi_{\text{c}}}{\lambda_{\text{c}}} \right)$. Now, $\frac{dH(r_{\text{n}})}{d r_{\text{n}}} = e^{-\lambda_{\text{n}} r_{\text{n}}} \left( - \pi_{\text{c}}  -\lambda_{\text{n}} \pi_{\text{n}}(r_{\text{n}} - \kappa) \right) + \pi_{\text{c}} e^{-\lambda_{\text{c}} r_{\text{c}}}$. Setting this to zero results in the fixed-point equation given by~\eqref{fixed_point}.
\end{proof}
\end{lemma}

\begin{lemma} \label{star_lemma}
A unique fixed-point solution exists for the optimal rebate for non-MC applications given by~\eqref{fixed_point} only if $\pi_{\text{c}} \leq \frac{\lambda_{\text{n}}}{2 \lambda_{\text{n}} - \lambda_{\text{c}}}$ and is expressed as follows:
\begin{align}
r_{\text{n}}^* = \log \left( \frac{\lambda_{\text{n}} \pi_{\text{n}}}{ \pi_{\text{c}}(\lambda_{\text{n}} - \lambda_{\text{c}})  }\right).
\end{align}
\begin{proof}
Let $L(r_{\text{n}}) = \kappa + \frac{ \pi_{\text{c}} ( e^{ r_{\text{n}}(\lambda_{\text{n}} - \lambda_{\text{c}})}   - 1   ) }{\pi_{\text{n}} \lambda_{\text{n}}}$. Since $L(r_{\text{n}})$ is an exponentially increasing function of $r_{\text{n}}$, so a unique fixed-point only exists if $\frac{dL(r_{\text{n}}^*)}{dr_{\text{n}}} = 1$. Solving this results in the expression for optimal rebate for non-MC applications given by the lemma. Furthermore, a valid solution only exists if $r_{\text{n}} \geq 0$ which implies that $\frac{\lambda_{\text{n}} \pi_{\text{n}}}{ \pi_{\text{c}}(\lambda_{\text{n}} - \lambda_{\text{c}}) } \geq 1$ resulting in the condition provided for $\pi_{\text{c}}$ in the lemma.
\end{proof}
\end{lemma}

Now, for optimality of $r_{\text{c}}^*$ and $r_{\text{n}}^*$, it is sufficient to show that the solution to the relaxed problem satisfies the IC$_{\text{n,c}}$ constraint as well. Since the IC$_\text{c,n}$ constraints bind with equality, we know that
\begin{align}\label{eq1}
p_{\text{n}} - p_{\text{c}} = r_{\text{n}} - r_{\text{c}} + \frac{e^{-\lambda_{\text{c}} r_{\text{n}}}}{\lambda_{\text{c}}} - \frac{e^{-\lambda_{\text{c}} r_{\text{c}}}}{\lambda_{\text{c}}}.
\end{align}
In order for IC$_{\text{n,c}}$ constraint to be specified, we need to show the following:
\begin{align}\label{eq2}
r_{\text{n}} - r_{\text{c}} - (p_{\text{n}} - p_{\text{c}} )   + \frac{e^{-\lambda_{\text{n}} r_{\text{n}}}}{\lambda_{\text{n}}}   - \frac{e^{-\lambda_{\text{n}} r_{\text{c}}}}{\lambda_{\text{n}}}  \geq 0.
\end{align}
Substituting \eqref{eq1} into \eqref{eq2} results in the following:
\begin{align}
\frac{e^{-\lambda_{\text{n}} r_{\text{n}}} }{\lambda_{\text{n}}}  - \frac{e^{-\lambda_{\text{n}}r_{\text{c}} } }{\lambda_{\text{n}}} + \left( \frac{e^{-\lambda_{\text{c}}r_{\text{c}}} }{\lambda_{\text{c}}} - \frac{e^{-\lambda_{\text{c}}r_{\text{n}}} }{\lambda_{\text{c}}} \right) \geq 0,
\end{align}
which can be shown to be true for $r_{\text{c}}^* = \kappa$ and $r_{\text{n}}^*$ given by Lemma~\ref{star_lemma} using the Taylor series expansion.
Once the optimal rebates have been obtained, the optimal advance payment can be obtained using the binding constraints IR$_{\text{n}}$ and IC$_{\text{c,n}}$ as expressed by the following corollary.
\begin{corollary}\label{advance_payment_lemma}
In the optimal contract offered to the IoT applications, the advance payment for reserving a TF block by MC and non-MC applications that maximizes the expected profit of the operator can be obtained as follows:
\begin{align}
p_{\text{n}}^* = r_{\text{n}}^* + \frac{1}{\lambda_{\text{n}}} e^{-\lambda_{\text{n}}r_{\text{n}}^*},
\end{align}

\begin{align}
p_{\text{c}}^* = p_{\text{n}}^* + r_{\text{c}}^* - r_{\text{n}}^* + \frac{1}{\lambda_{\text{c}}} \left( e^{-\lambda_{\text{c}} r_{\text{c}}^*}  -  e^{-\lambda_{\text{c}} r_{\text{n}}^* } \right).
\end{align}
\end{corollary}

This completes the solution to the optimal contract design for spectrum reservation in IoT systems.

\begin{figure}
  \centering
  \includegraphics[width=3.5in]{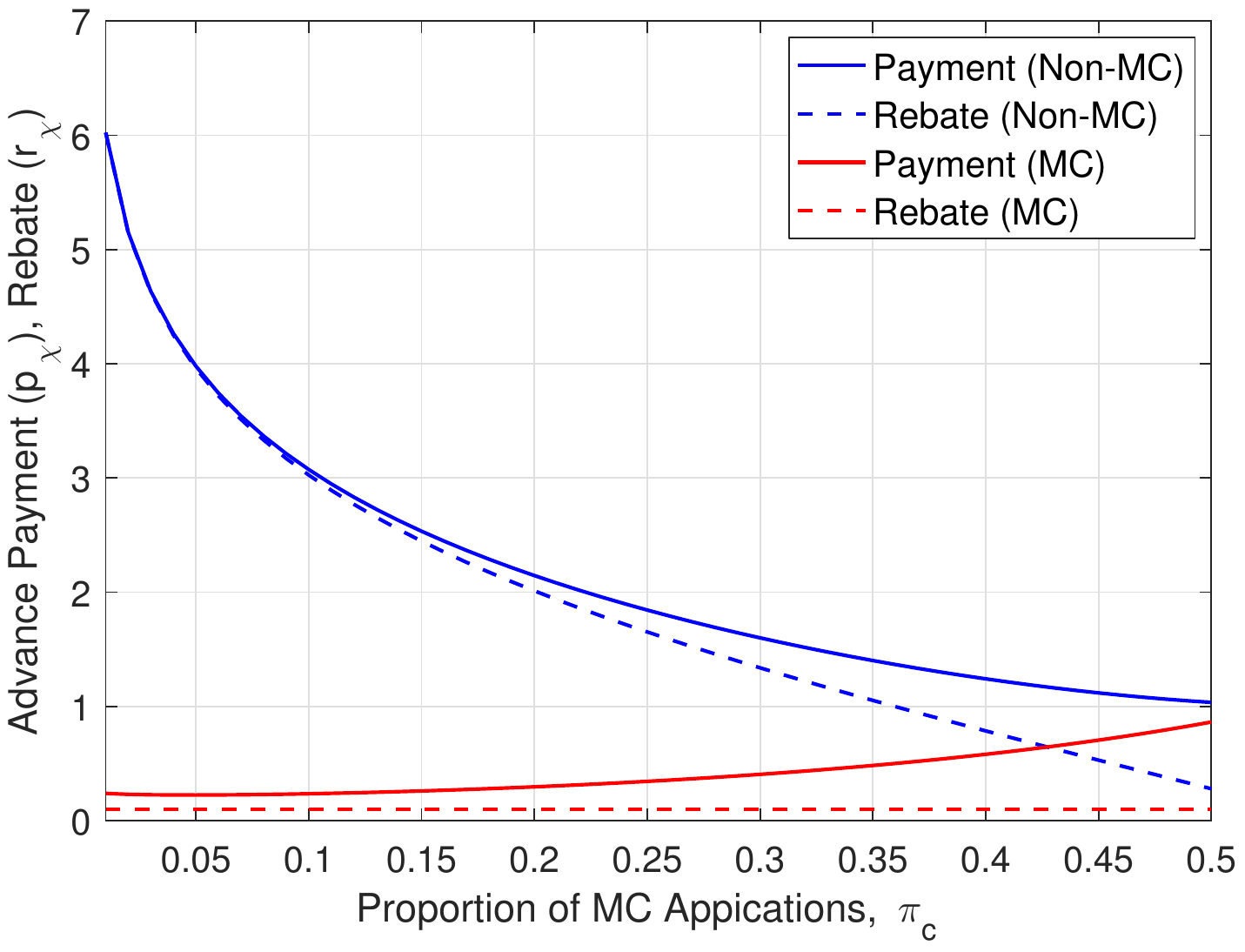}\\
  \caption{Effect of varying proportion of MC applications in the network.\vspace{-0.0in}}\label{Fig:Result_pi_c}
\end{figure}

\vspace{-0.0in}
\section{Results}

In this section, we present the results obtained from analyzing the designed optimal spectrum reservation menu of contracts for varying system parameters. We first describe the network setup and the assumed system parameters. We assume a single monopoly IoT network operator and consider the scenario where an access point coordinates the communication between various IoT devices and the internet. The AP enforces a reservation based spectrum access system where the applications can only communicate with the AP in a particular TF block if they have a prior reservation. Furthermore, the AP does not know about the nature of the application using the spectrum. Therefore, it only offers two contracts to the applications based on prior information about the proportions of the application types and the distribution of their spectrum reservation utility.

The system parameters are selected as follows: We assume that 20\% of the applications connected to the AP are MC and the rest are non-MC, i.e., $\pi_{\text{c}} = 0.2$ and $\pi_{\text{n}} = 0.8$ unless otherwise stated. The time average of the cost of obtaining an idle channel to the network operator is considered to be $\kappa = 0.1$ monetary units (MU). Unless otherwise stated, we will use $\lambda_{\text{c}} = 0.2$ MU$^{-1}$ and $\lambda_{\text{n}} = 1$ MU$^{-1}$ implying that the expected utility of the MC applications, i.e., 5 MU, is higher than the non-MC applications, i.e., $1$ MU before the true utility is learnt. Note that the parameter selection is made for illustrative purposes and does not affect the generality of the results. Using the selected parameters, the optimal amount of rebate offered to the applications is obtained using Lemma~\ref{rebate_lemma} and Lemma~\ref{star_lemma}. Consequently, the optimal advance payments can be obtained using results in Corollary~\ref{advance_payment_lemma}. The main observations are described in the sequel.

\begin{figure}
  \centering
  \includegraphics[width=3.5in]{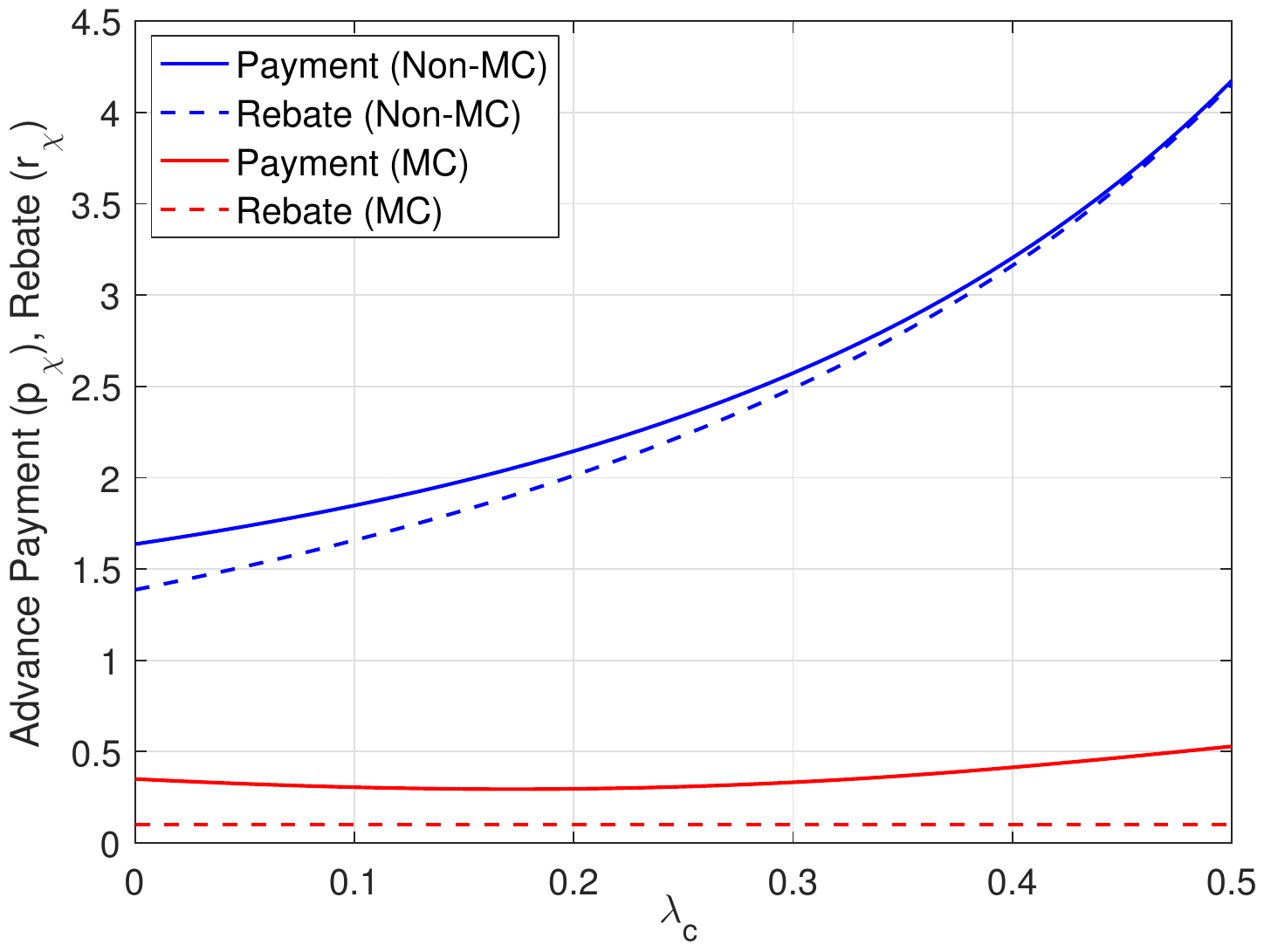}\\
  \caption{Effect of decreasing average valuation of MC applications.\vspace{-0.0in}}\label{Fig:Result_lambda_c}
\end{figure}

In Fig.~\ref{Fig:Result_pi_c}, we plot the optimal menu of contracts against varying proportion of MC applications in the network. It can be observed that the optimal amount of rebates (shown by dotted lines) are always lower than the corresponding advance payments (shown by solid lines). Moreover, that the rebate to MC applications in the optimal contract is always equal to the average cost of obtaining a channel, i.e., $\kappa = 0.1$. It is observed that as the proportion of MC applications increases, the advance payment for MC applications in the optimal contract reduces while the optimal rebate for MC applications is fixed at $\kappa$. On the other hand, the proportion of non-MC applications decreases implicitly and this also results in a decrease in the advance payment and the rebate. Note that as the proportion of MC applications increase to 0.5, i.e., the MC and non-MC applications are equally distributed, the return profit margin (difference between advance payment and rebate) of the operator becomes equal for both menus.

Fig.~\ref{Fig:Result_lambda_c}, investigates the effect of the average utility of the MC applications prior to contracting on the optimal contracts. Note that increasing $\lambda_{\text{c}}$ implies that the expected utility of the MC applications decreases. We fix $\lambda_{\text{n}} = 1$ MU$^{-1}$ and sweep $\lambda_{\text{c}}$ to $\lambda_{\text{c}}=0.5$. Note that as $\lambda_{\text{c}}$ increases, i.e., the expected utility decreases, the payment and rebate amounts increase in general to make up for the revenue. Furthermore, the increase is proportional to the composition of the application types.

\begin{figure}
  \centering
  \includegraphics[width=3.5in]{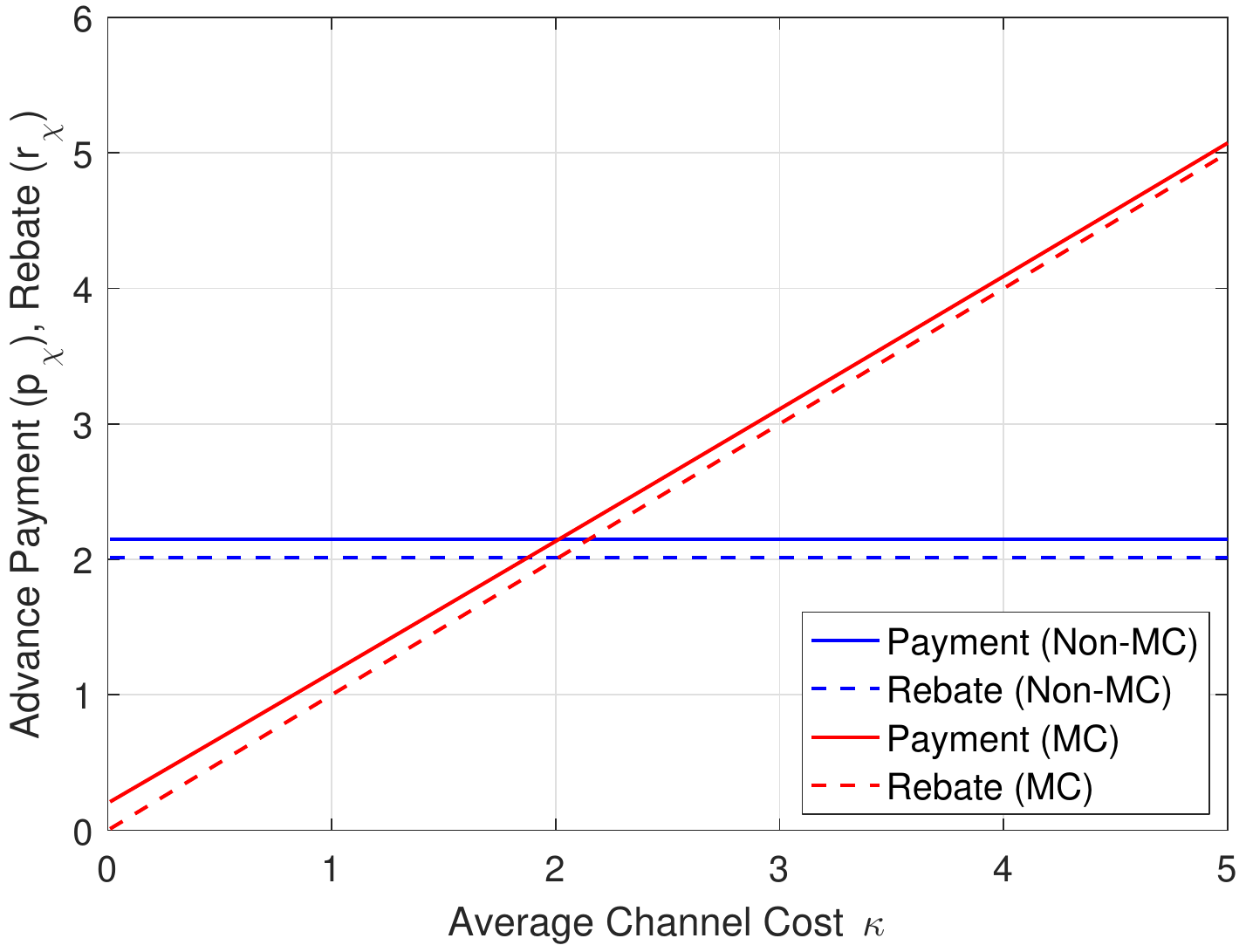}\\
  \caption{Effect of average channel cost on the reservation contract.\vspace{-0.0in}}\label{Fig:Result_kappa}
\end{figure}

Finally, in Fig.~\ref{Fig:Result_kappa}, we investigate the effect of varying the average cost of opportunistically obtaining an idle TF block in the licensed spectrum. It is intuitive that the advance payments and the rebates for MC applications increase with the increasing cost. However, it is interesting to note that the rebates increase linearly with the cost while the advance payments increase non-linearly. As the cost becomes significantly high, the rebate policy in the optimal dynamic contract effectively becomes a full refund policy. In other words, the advance payment becomes high on the agreement that a full refund will be issued in case of cancellation later. Furthermore, the advance payments and the rebates designed for non-MC applications is independent of the cost of obtaining the channel.

\section{Conclusion \& Future Work}
In this paper, we present a dynamic mechanism design framework to establish a spectrum reservation contract for mission-critical IoT devices in the licensed spectrum using ultra narrow band technology where time-frequency blocks can be reserved ahead of time. If the the channel, is not required later after the initial reservation, the service provider pays back a part of the reservation payment as a rebate. We develop optimal dynamic contracts where the applications learn the reservation utility sequentially over time. The optimal advanced payments and the rebate amounts are obtained in closed form that maximize the expected profit of the network operator. Finally, the behaviour and properties of the contract terms is analyzed against different system parameters.

In the current work, we have restricted our design and analysis to use a broad classification of applications, i.e., either MC or non-MC types. In practice, there can be several categories of applications which have different levels of sensitivities to spectrum access delay. Therefore, future extensions in this direction can further enhance the developed model to a continuum of application types covering various levels of mission criticality which may offer higher price discrimination and consequently higher profitability for the operators.

\bibliographystyle{IEEEtran}
\bibliography{icc_references}

\end{document}